# BARS: a Blockchain-based Anonymous Reputation System for Trust Management in VANETs

Zhaojun Lu*, Qian Wang†, Gang Qu†, and Zhenglin Liu*
* School of Optical and Electronic Information, Huazhong University of Science and Technology, Wuhan, China
† Department of Electrical and Computer Engineering, University of Maryland, College Park, United States
E-mail: {D201377521, liuzhenglin}@hust.edu.cn, {qwang126, gangqu}@umd.edu

*Abstract*—The public key infrastructure (PKI) based authentication protocol provides the basic security services for vehicular ad-hoc networks (VANETs). However, trust and privacy are still open issues due to the unique characteristics of vehicles. It is crucial for VANETs to prevent internal vehicles from broadcasting forged messages while simultaneously protecting the privacy of each vehicle against tracking attacks. In this paper, we propose a blockchain-based anonymous reputation system (BARS) to break the linkability between real identities and public keys to preserve privacy. The certificate and revocation transparency is implemented efficiently using two blockchains. We design a trust model to improve the trustworthiness of messages relying on the reputation of the sender based on both direct historical interactions and indirect opinions about the sender. Experiments are conducted to evaluate BARS in terms of security and performance and the results show that BARS is able to establish distributed trust management, while protecting the privacy of vehicles.

*Keywords—vehicular ad-hoc networks, blockchain, trust management, reputation system, privacy*

## I. INTRODUCTION

It is estimated that the number of registered vehicles will reach 2 billion within the next 10 to 20 years [1]. Recently, vehicular ad-hoc networks (VANETs) have been suggested as the foundation of intelligent transportation systems (ITSs) to improve the transportation efficiency and ensure the safety of both vehicles and pedestrians. Two types of communications, namely vehicle-to-vehicle (V2V) and vehicles-to-infrastructure (V2I) communication [2] are established in VANETs to promote cooperation among vehicles and to share valuable driving information. Through dedicated short range communication (DSRC) radio, vehicles exchange messages with nearby vehicles in V2V and communicate directly with roadside units (RSUs) in V2I [2].

However, the unique characteristics of VANETs such as high mobility and volatility make it vulnerable to various kinds of attacks. Security, privacy, and trust should be taken into account from the beginning stages of designing VANETs. Although major security services have been well-studied in other fields that can provide secure communication channels against external attackers, trust management and privacy protection for vehicles are still open issues for VANETs. Specifically, it is fairly difficult to deal with misbehaviors and distribution of forged messages from authenticated vehicles. These forged messages could not only decrease the transportation efficiency but also in the worst cases, cause accidental events that can threaten human life [3]. In addition, internal attackers can easily track other vehicles or profile the drivers' actions by analyzing all the broadcasted messages in VANETs.

In order to provide a trust communication environment, trust management evaluates the trustworthiness of messages based on both direct historical interactions and indirect opinions about the senders [4]. An effective trust model should have the following properties:

**Efficiency**. It should be efficient in determining the trustworthiness of a warning message in both congested and sparse situations.

**Privacy**. It should not reveal any sensitive information such as the real identity of the senders.

**Robustness**. It should be resistant against attacks aiming at deceiving the trustworthiness evaluation or disabling the trust model.

The blockchain is the underlying technology of the Bitcoin protocol that emerged in 2008 [5]. It is a distributed public ledger encrypted using Merkel tree and hash function and has a consensus mechanism based on a proof of work (PoW) algorithm. These significant features of blockchain make it potential for constructing the desirable trust model in VANETs. All the broadcasted messages and actions of vehicles will be written into the immutable and unforgeable record, which can be verified and audited by every entity in the network. However, the transparency of blockchain means privacy is not considered naturally. By reviewing the ledger, the actions made with any public key is traceable to a real identity.

In this paper, we propose a blockchain-based anonymous reputation system (BARS) to establish distributed trust management while simultaneously protecting the privacy of vehicles. The main contributions of BARS are twofold:

First, we exploit the features of blockchain to extend conventional public key infrastructure (PKI) with an efficient privacy-preserving authentication mechanism. The linkability between the public key and the real identity of a vehicle is eliminated when a certificate authority (CA) operates the certificate issuance and revocation. All the actions of CA are recorded in blockchain transparently without revealing sensitive information about vehicles so that the public key can be used as an authenticated pseudonym. The law enforcement authority (LEA) is responsible for managing BARS and recording the pairs of public key and real identity in case of disputes.

Second, we design the reputation management algorithm to evaluate the trustworthiness of each vehicle according to the authenticity of broadcasted messages as well as opinions from other vehicles. All the messages are recorded in blockchain as persistent evidence for LEA to evaluate the reputation score for each vehicle. The reputation score provides an incentive for internal vehicles to prevent misbehaviors and mitigate the distribution of forged messages.

The remainder of this paper is organized as follows: Sections II surveys the existing trust models for VANETs. Section III introduces the background knowledge, including the rationale of blockchain, certificate transparency, the components of BARS, and the necessary assumptions. Section IV proposes the anonymous authentication to protect the privacy of vehicles. The reputation evaluation algorithm is presented in Section V. Finally, we conduct simulations in Section VI to evaluate BARS in terms of security and performance and give the conclusion.

## II. TRUST MODELS IN VANETs

As illustrated in Fig. 1, state-of-the-art trust models can be classified into three categories: (1) entity-centric trust models, (2) data-centric trust models, and (3) combined trust models. None of them meets all the requirements of a desirable trust model for VANETs.

### A. Entity-centric Trust Models

Entity-centric trust models focus on evaluating the trustworthiness of vehicles. The main methods to achieve this efficiently and accurately are to establish a reputation system or to make a decision according to the opinions of neighbors. There are several typical works. Minhas et al. [6] develops a multifaceted trust modeling approach to detect the entities that are generating malicious data. This method incorporates role-, experience-, priority-, and majority-based trust to make a real-time decision. Mármol at al. [7] proposes a trust and reputation infrastructure-based proposal (TRIP) for VANETs to quickly and accurately distinguish malicious or selfish nodes with the help of RSUs. Haddadou et al. [8] propose a distributed trust model (DTM2) to allocate credits to nodes and securely manage these credits. Due to the high mobility of vehicles, it is difficult to collect enough information to calculate the reputation score of a specific node. Moreover, another serious issue that has not been resolved is how to ensure the security of the reputation system itself.

### B. Data-centric Trust Models

Data-centric trust models focus on the trustworthiness of received data. In order to verify the trustworthiness of the received data accurately, the models need cooperative information from various sources such as neighbor vehicles or RSUs. Gurung et al. [9] propose a trust model to directly evaluate the trustworthiness of a message based on various factors such as content similarity, content conflict, and route similarity. Huang et al. [10] develop a voting system with different voting weights according to its distance from the event. Rawat et al. [11] propose a deterministic approach to measure the trust level of the received message by using received signal strength (RSS) for distance calculations as well as the vehicle's geolocation (position coordinate). Hussain et al. [12] suggest email-based social trust and social networks-based trust to establish and manage data level trust. The major drawbacks of data-centric trust models are latency and data sparsity. Respectively, large numbers of data from various sources may contain redundant information, which will increase latency or overwhelm the significant information. On the contrary, data sparsity is prevalent in VANETs. It is unrealistic for data-centric trust models model to perform well without enough information.

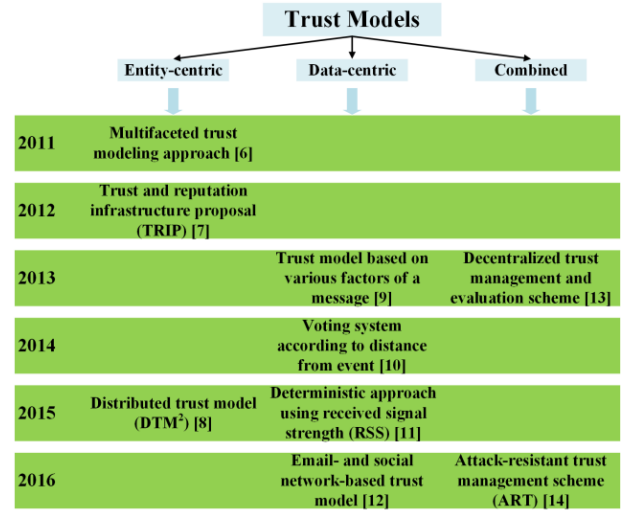

Fig. 1. Trust models proposed in recent years.

### C. Combined Trust Models

Both entity and data are the main objects in this category. Combined trust models not only evaluate the trust level of vehicles but also calculate the trustworthiness of data [13]. Thus, these models inherit the benefits and drawbacks of entity-centric and data-centric trust models. The attack-resistant trust management scheme (ART) proposed by Li et al. [14] cope with malicious attacks in VANETs. Trustworthiness of data is evaluated based on the received data from multiple vehicles. Trustworthiness of a node is determined based on functional trust and recommendation trust, which respectively indicate whether a node can fulfill its functionality and what the trust level of the recommendations from it is. The proposed scheme does not take into account data sparsity, which is pervasive in VANETs.

## III. BACKGROUND OF BARS

Before elaborating how BARS works, we will first introduce the way blockchain works and how it guarantees security. Then we will point out the problems of conventional PKI and explain the concept of certificate transparency. Finally, we will introduce the major components of BARS with their functions and give some necessary assumptions.

## A. Blockchain

The blockchain is a computational paradigm which emerged with the Bitcoin protocol in 2008 [5]. It is a distributed ledger containing all transactions ever executed within the network. The ledger is enforced with cryptography and carried out collectively in a peer-to-peer network. As a secure and decentralized computational infrastructure, it is widely acknowledged as a disruptive solution for the problems of centralization, privacy and security when storing, tracking, monitoring, managing and sharing data [15].

## B. Certificate Transparency

Certificate transparency [16] is invented by Google and aims to prevent transport layer security (TLS) certificate authorities from issuing public key certificates for a domain without it being visible to the owner of the domain. This technology is being built into Google Chrome aiming at website certificates.

The core idea of certificate transparency is that a public append-only log is maintained to show all the certificates that have been issued. Anyone can append a certificate to the log. Auditors can obtain two types of proofs: (a) a proof that the log contains a given certificate and (b) a proof that a snapshot of the log is an extension of another snapshot (i.e., only appends have taken place between the two snapshot).

## C. Components of BARS

**Certificate Authority (CA)**. CA issues and revokes certificates only if it gets a warrant from LEA. All the actions of CA will be recorded transparently in the blockchain and can be verified by every entity in the VANETs

**Law Enforcement Authority (LEA)**. The functions of LEA include registration, monitoring vehicles, and evaluating the reputation scores of each vehicle. LEA authorizes CA for certificates issuance and revocation. LEA also keeps the database that contains correlation between vehicles' public keys and the real identities with high-level security.

**Certificate**. The certificate contains the expiration date, the public key, and the reputation score but no real identity so that the privacy of the vehicle is protected.

**Blockchain for certificates (CerBC)**. CerBC acts as the public ledger for all the issued certificates. It provides efficient proof of presence for received certificates.

**Blockchain for revoked public keys (RevBC)**. RevBC acts as the public ledger for all revoked public keys. It provides efficient proof of absence for the sender's public key.

**Blockchain for messages (MesBC)**. All the broadcasted messages will be recorded in MesBC as persistent evidence in case of disputes.

**Roadside Unit (RSU)**. The global consensus is based on the proof of work (PoW) provided by RSUs. As long as more than half of the RSUs are not compromised, the security of BARS can be guaranteed.

**Vehicle**. On one hand, vehicles can monitor CA and LEA by verifying that all the messages are recorded in blockchains. On the other hand, vehicles can monitor each other to prevent misbehaviors and forged messages.

## D. Assumptions

First, the asymmetric cryptography in PKI is able to provide a secure communication channel between entities as long as the secret key is not stolen.

Second, law enforcement authority (LEA) has enough security levels to keep the dataset containing the correlation between vehicles' public keys and the real identities safe.

Third, we assume that it is beyond the adversaries' capability to compromise more than half of the RSUs, which is the prerequisite to ensure the blockchain itself is secure.

## IV. ANONYMOUS AUTHENTICATION

Anonymous authentication is fundamental to trust communication and privacy protection. In BARS, CA and LEA are responsible for three major functions: system initialization, certificate update, and public key revocation. We will first respectively present the three functions and then explain the process of privacy-preserving authentication.

## A. System Initialization

Initially, each entity generates a pair of private and public keys. When vehicle A enters the network, it uses the secure channel to submit LEA its initial public key and materials to prove its legal identity. LEA will send a signed warrant to CA if the materials are valid. Next, CA will issue an initial certificate to vehicle A.

Note that the submitted material contains vehicle A's private information. Only LEA preserves them in the database with high-security level, which will be used for tracking the vehicle's real identity in case of disputes.

## B. Certificate Update

Vehicle A will send a request to LEA for a certificate update in the following situations: First, before the current certificate expires. Second, if the security of its private key is threatened. Third, if it requests to replace its public key for privacy consideration. The public key, reputation score, and expiration date are updated in a new certificate. Fig. 2 illustrates the steps to update certificate anonymously.

**Step 1.** Vehicle A generates a new pair of public key and private key $\{PU^n_A, PR^n_A\}$.

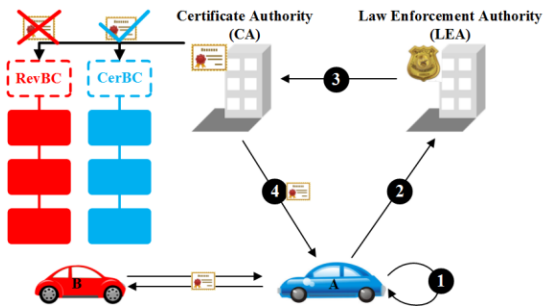

Fig. 2. Certificate update process.

**Step 2.** Vehicle A sends LEA certificate update request encrypted with LEA's public key $PU_{LEA}$. The request includes vehicle A's current public key $PU^{n-1}_A$, updated public key $PU^n_A$, proofs of real identity, and the signature $Sig_A$ using A's current private key $PR^{n-1}_A$.

**Step 3.** If vehicle A's request is verified, LEA will send CA a signed warrant. For the purpose of privacy protection, the linkability between A's current and updated public key is unknown by CA.

**Step 4.** CA will verify the signature in the warrant. Then, an updated certificate containing the updated public key $PU^n_A$, A's reputation score $Rpt_A$, and the expiration time $T_A$ will be issued to vehicle A publicly and recorded into CerBC.

$$C^n_A = \langle PU_{CA}, Sig_{CA}, PU_{LEA}, Sig_{LEA}, PU^n_A, Rpt_A, T_A \rangle$$

### C. Public Key Revocation

Vehicle A's public key should be revoked before its expiration date when vehicle A requests to replace its public key or A's misbehavior is discovered by LEA. In order to provide revocation transparency, LEA sends signed revocation instructions to CA that contain the revoked public key $PU_{rev}$ and the revocation time $T_{rev}$. Then CA broadcasts the revocation messages that contains the revoked public key, the timestamp, and signatures of CA and LEA:

$$Rev = \langle PU_{CA}, Sig_{CA}, PU_{LEA}, Sig_{LEA}, PU_{rev}, T_{rev} \rangle$$

The RSUs will verify all the revocation messages in a predefined interval, delete the expired public key, and lexicographically insert revoked public keys into RevBC. RevBC has a lexicographical Merkle tree [17] called LexTree that can provide efficient proof of absence.

### D. Authentication Process

Vehicle A's certificate $C_A$ is used for authentication. When vehicle B receives $C_A$, it first checks whether the certificate is expired. If not, B will look up the CerBC and RevBC to make sure $C_A$ is present in CerBC but $PU_A$ is absent in RevBC, which means $PU_A$ is issued and not revoked by CA. Then we will give the details of proof of presence and proof of absence. The secure analysis will be presented in Section VI.

*1) Proof of Presence:* Fig. 3 illustrates how to prove $C_4$ is present in CerBC. A tuple (dir, hash) is enough for the proof of presence for $C_4$, in which dir = {left, left, right} and hash = {$h_3$, $h_{12}$, $h_{56}$}. The receiver can get the root hash value using the tuple. If this root hash value is equal to the root recorded in CerBC, it means $C_4$ and the associated public key is valid.

*2) Proof of Absence:* It is unreasonable to force a vehicle to demonstrate that its public key is revoked. Thus, a proof of absence is necessary for a vehicle to convince others its public key is valid. As shown in Fig. 4, all the revoked public keys (not expired) are recorded in the data structure based on lexicographical Merkle tree. Vehicle A should prove that two adjacent public keys ($PU_7$, $PU_8$) exist in the left-right traversal of the tree meanwhile $PU_7 \leqslant PU_B \leqslant PU_8$ lexicographically. A tuple (PU, hash) is used for proof of absence, in which PU = {$PU_7$, $PU_8$, $PU_{10}$, $PU_6$} and hash = {$h_9$, $h_{12}$, $h_4$}. Similarly, if the hash value calculated using the tuple is same as $h_6$ that is recorded in RevBC, it means that A's public key is absent in RevBC.

## V. REPUTATION MANAGEMENT

Blockchain-based anonymous reputation system (BARS) relies on the reputation score of a vehicle to determine the trust level of broadcasted messages. In this section, we will elaborate how BARS can provide a trust communication environment while protecting the privacy of vehicles.

### A. Different Types of Messages

There are three types of messages: beacon messages, alert messages, and disclosure messages. Periodically, vehicles broadcast beacon messages containing driving status for traffic management. Alert messages will be broadcasted when an emergency happens, including hard braking or losing control. If any vehicle disputes the authenticity of the received message or witnesses misbehaviors, they can send disclosure messages to LEA. Next, LEA will make a judgment that affects the reputation scores of related vehicles. According to the criticality of the emergency, alert messages have three levels.

**Level 1**. When vehicle A loses control, it will broadcast level 1 alert message to avoid collision automatically.

**Level 2**. Level 2 alert message is used for forewarning nearby vehicles before the sender changes its driving status, including braking, lane changing, etc.

**Level 3**. In case of poor road conditions such as obstruction or road damage, passing vehicles will broadcast level 3 alert messages to alert vehicles behind to keep caution.

### B. Reputation Evaluation Algorithm

The reputation evaluation algorithm consists of a reward mechanism and a punishment mechanism. There are two kinds of actions that will be rewarded. First, vehicle A broadcasts alter messages honestly and actively. Second, vehicle A sends disclosure messages to LEA when A witnesses misbehaviors or receives forged messages. On the contrary, there are also two kinds of actions which will be punished. First, vehicle A is disclosed for misbehaviors or broadcasting forged messages. Second, vehicle A abuses disclosure messages to slander other vehicles.

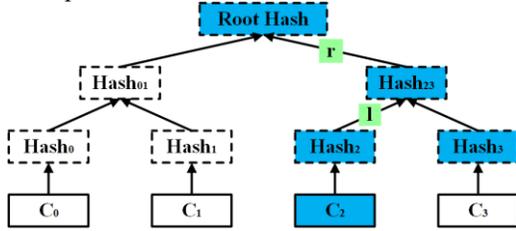

Fig. 3. Proof of presence.

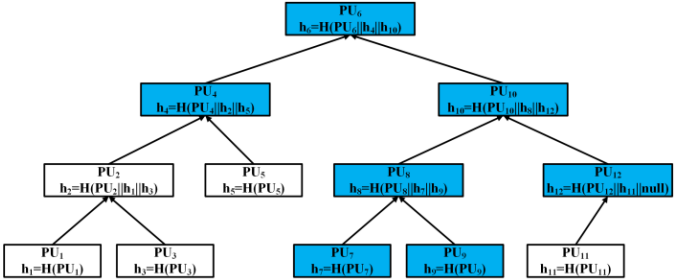

Fig. 4. Proof of absence [17].

There are several factors affecting the evaluation of reputation scores as follows:

$L$: The level of alert messages, $L = 1, 2, 3$.

$D_r$: The relative density of vehicles, $D_r = D/D_{aver}$. In this paper, $D_{aver}$ is set to 20 vehicles per Km.

$S$: The sequence of the senders, $S = 0, 1, \ldots, n$. $S$ of the first vehicle to broadcast an alert message will be set to 0.

In addition, we set reward coefficient $\alpha$ and penalty coefficient $\beta$ to implement reward mechanism and penalty mechanism:

$$R(L, S, D_r) = \alpha * D_r * \frac{1}{e^S * L} \qquad P(L, S, D_r) = (-1) * \beta * D_r * \frac{1}{e^S * L}$$

As illustrated in Algorithm 1, if no receiver disputes the authenticity of an alert message, the reputation score will increase based on reward mechanism. On the contrary, if any receivers send disclosure messages to dispute the authenticity of the alert message, LEA will collect evidence to make a judgment. The vehicles who broadcast forged alert messages will be punished heavily, while also publishing the vehicles who abuse disclosure messages.

**Algorithm 1** Reputation Evaluation Algorithm

**Require:** $M_A$: Alert message broadcasted by vehicle $V_i (i = 1, 2...n)$; $M_D$: Disclosure message broadcasted by vehicle $V_j (j = 0, 1...m)$; $R'_i, R'_j$: Current reputation score of $V_i$ and $V_j$; $S_i, S_j$: The sequence of the senders; $D_r$: The relative traffic density.
**Ensure:** $R_i, R_j$: Updated reputation of $V_i$ and $V_j$.
1: **if** $j = 0$ **then**
2:   **for** each $V_i$ **do**
3:     $R_i \leftarrow R'_i + (100 - R'_i) \cdot R(M_A.L, S_i, D_r)$
4:   **end for**
5: **else**
6:   **if** $M_A$ is authentic **then**
7:     **for** each $V_i$ **do**
8:       $R_i \leftarrow R'_i + (100 - R'_i) \cdot R(M_A.L, S_i, D_r)$
9:     **end for**
10:     **for** each $V_j$ **do**
11:       $R_j \leftarrow R'_j + 25 \cdot P(M_A.L, S_j, D_r)$
12:     **end for**
13:   **else**
14:     **for** each $V_i$ **do**
15:       $R_i \leftarrow R'_i \cdot (1 + P(M_A.L, S_j, D_r))$
16:     **end for**
17:     **for** each $V_j$ **do**
18:       $R_j \leftarrow R'_j + 50 \cdot R(M_A.L, S_j, D_r)$
19:     **end for**
20:   **end if**
21: **end if**
22: **return** $R_i, R_j$

## VI. RESULTS AND ANALYSIS

### A. Security Analysis

*1) Security of Certificates:* The nature of blockchain satisfies the security requirement of certificates. RSUs verify the signatures of CA and LEA in the certificate issuance or revocation messages, and record them into CerBC and RevBC respectively. The global consensus is provided by the PoW of RSUs to guarantee that each vehicle has an identical public ledger that consists of authenticated certificates and revoked public keys.

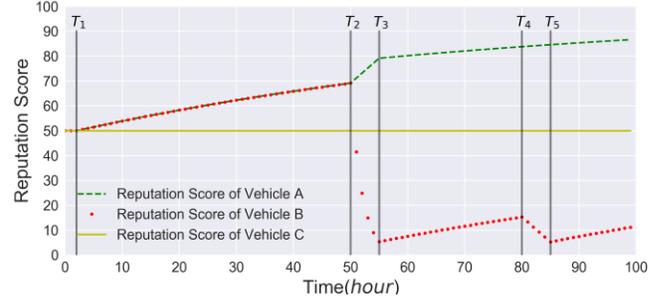

Fig. 5. Reputation score of three vehicles.

*2) Security of Broadcasted Messages:* Proof of presence in CerBC and proof of absence in RevBC ensure the authentication of the public key of vehicle A. Then, vehicle A will use its private key $PR_A$ to generate a signature for each broadcasted message and receivers can use A's public key $PU_A$ to verify the signature. RSUs and vehicles cooperatively recorded all the broadcasted messages into the chronological MesBC, which is the persistent evidence when a dispute happens.

*3) Privacy of Vehicles:* Vehicle A uses public key as the pseudonym to break the linkability between the public key and the real identity. For the trade-off between security and privacy, the database of identity-public key pairs is stored with high-security level in LEA. This means only LEA knows the real identity of any public key so that LEA is able to track the malicious vehicle when it performs misbehaviors or broadcasts forged messages.

### B. Validity of Reputation Evaluation Algorithm

We consider three vehicles who perform different behaviors in 100 hours. As Fig. 5 presents, from $T_1$ to $T_2$ and $T_3$ to $T_4$, vehicle A and B actively broadcast authentic messages and their reputation scores increase. From $T_2$ to $T_3$, vehicle B broadcasts five forged messages and is disclosed by A. Thus, A's reputation score increases whereas B gets punished. From $T_4$ to $T_5$, vehicle B abuses five disclosure messages to slander other vehicles. As a result, B's reputation score decreases. Vehicle C refuses to participate in BARS. The results show that the reputation score effectively reflects vehicle's behavior.

### C. Performance Evaluation

All the experiments are conducted using 2.5 GHz Intel Core i5 and 8 GB 1600 MHz DDR3. The time consumption to compute an SHA-256 is less than $t_1 = 0.01$ ms per 1 KB of input.

*1) Storage & Transmission Overhead:* A block header would be about 80 bytes [5]. Suppose that blocks are generated every 10 minutes, the storage overhead for one blockchain is 80 bytes * 6 * 24 * 365 = 4.2 MB per year. An authentication packet for a public key consists of the associated certificate (about 100 bytes), the tuple for proof of presence in CerBC, and the tuple for proof of absence in RevBC. Suppose there are n issued certificates and m revoked public keys, the total storage overhead of an authentication packet is S = 100 bytes + 32 bytes * $\log_2^n$ + (32 bytes + 8 bytes) * $\log_2^m$. The main source of transmission overhead is the authentication packets and beacon messages (about 100 bytes for each). Suppose that vehicle A receives i authentication packets per second from nearby vehicles, in which there are j new public keys. A will automatically discard the rest of an authentication packet if the public key is in the list of authenticated vehicles. Thus, the total transmission overhead is Tran = 100 bytes * (i - j) + S bytes * j per second. If we set i = 100, j = i * 10%, n = 1 000 000, m = n * 10%, the total transmission overhead is about 0.177 M/s. The result shows that the storage & transmission overhead of anonymous authentication is acceptable.

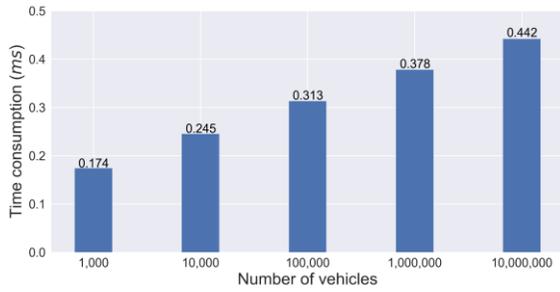

Fig. 6. Time consumption of authentication.

*2) Computation Overhead:* The proof of presence and proof of absence are based on SHA-256 and can be done in time and space $O(\log^n)$ (n is the number of elements in the Merkle tree). Theoretically, the time consumption to authenticate one public key is T = $t_1$ * ( $\log_2^n$ + $\log_2^m$ ). We assume that m = n * 10%, i.e. 10% of the public keys are revoked. Fig. 6 illustrates the time consumption of anonymous authentication in different scales of VANETs. The logarithmic-size proofs of presence and absence provide efficient authentication in large-scale network.

## VII. CONCLUSION

In this paper, we address the issues of trust and privacy in VANETs. In order to prevent the distribution of forged messages from authenticated vehicles while protecting the identity privacy of vehicles, a blockchain-based anonymous reputation system (BARS) is proposed for anonymous authentication and trust communication for VANETs. Vehicles use two blockchains (CerBC and RevBC) for authentication, which is based on proofs of presence and absence. Additionally, public keys act as the pseudonyms for anonymous communication and the linkability between the real identity and the public key is eliminated to protect vehicles' privacy. On the other hand, all the broadcasted messages are recorded in MesBC as persistent evidence to evaluate each vehicle's reputation. A reputation management algorithm is designed for trust communication to prevent the spread of forged messages and incentivize vehicles to expose misbehaviors. Finally, we analyze the security and validity of BARS and evaluate the performance. The results show that BARS effectively improves the trustworthiness of broadcasted messages and protects vehicle privacy with high efficiency.

ACKNOWLEDGEMENT

This work is supported in part of National Science Foundation of China (Grant No. 61376026 and No.61176026).